\documentstyle[preprint,aps]{revtex}

\begin{document}

\tighten

\def\ONE{\hbox{{\rm 1 \kern -.65em \rm I}}}
\def\R{{{\rm I} \kern -.15em {\rm R}}}
\def\Z{{{\rm Z} \kern -.35em {\rm Z}}}

\title{Representations of the Heisenberg algebra by difference operators}

\author{Andrzej Z. G\'orski\thanks{Address: Inst. of Nuclear Physics,
Radzikowskiego 134, 30--432 Krak\'ow, Poland, E--mail:
gorski@alf.ifj.edu.pl}
}
\address{Institute of Nuclear Physics, Cracow, Poland}

\author{Jacek Szmigielski\thanks{Address: Department of Mathematics and
Statistics, University of Saskatchewan, Saskatoon, Saskatchewan S7N 5E6,
Canada, E--mail: szmigiel@math.usask.ca}
}
\address{Dept. of Mathematics and Statistics, University of
Saskatchewan, Saskatoon, Canada}

\address{(August 19, 1998)}

\maketitle

\begin{abstract}
 We construct a class of representations of the Heisenberg algebra
in terms of the complex shift operators subject to
the proper continuous limit imposed by the correspondence principle.
We find a suitable Hilbert space formulation of our construction
for two types of shifts:
(1) real shifts, (2) purely imaginary
shifts. The representations involving imaginary shifts are free
of spectrum doubling.  We determine the corresponding coordinate and momentum
operators satisfying the canonical commutation relations.
The eigenvalues of the coordinate operator are in both cases discrete.
\end{abstract}

\newpage

{\it 1. Introduction.}
 The aim of this paper is to find a possibly wide class of
difference operator representations for the
coordinate and momentum operators leading to a generalization of the standard
quantum mechanics (QM) in a way compatible with the discreteness of
space--time.
 It is important to stress that at this stage we are at the level of
{\em kinematics}. Thus, we are looking for proper definitions of
the coordinate and momentum operators, their domains, the "classical"
limit" (to recover the standard QM in the continuous limit), the Hilbert
space etc.  We assume that the Heisenberg commutation
relations (CR) hold on some dense set of states.
 A deeper understanding of, seemingly, exotic realizations of
CR for which one of the operators is bounded might be needed to
advance this program.  In other words, the lack of such understanding
was perhaps responsible for the relatively unsuccessful, albeit very interesting, results
of the early attempts in this direction
\cite{Snyder,Pais,Das,Cole}.

 Our research is motivated by an old problem of the nature of the
space--time geometry below the Planck scale.  There have been many
speculations and indications that at this level new discrete
structures are likely to emerge (see the arguments put forward in
string theory \cite{KlebanovSuskind}, in various approaches to quantum
gravity \cite{Horowitz,Smolin,Bekenstein,Mazur}, in 't Hooft's
approach \cite{tHooft} {\it etc.})  and that, as a consequence, some
"unusual" representations of the Heisenberg algebra maybe appropriate
\cite{Douglas}.  Finally, discrete models are widely used because of
technical reasons, perhaps the most obvious examples beeing lattice
gauge theories \cite{Kogut} and Regge calculus approach to general
relativity \cite{Regge} to name only a few.

 The first step to realize this program has been taken in \cite{JMP},
where representations of the Heisenberg CR in terms of real shift
operators have been analyzed.  We have discussed the Bargmann--Fock
(annihilation--creation operators) as well as the Schr\"odinger
(position and momentum operators) representations.  In particular, we
have found that our coordinate operator has a discrete spectrum.  In
this paper we generalize our previous results by including the complex
shift operators as well. In particular, we find that the operators
discussed in \cite{Frappat,Mazur,Berezin} are included in our
scheme. Those operators correspond to purely imaginary shifts and in
the last two references they are used in a physically interesting
context of the black hole's energy quantization.

We work directly with the coordinate (Schr\"odinger) representation, as
opposed to the annihilation-creation type, for the former seems to be
physically more fundamental; the sentiment captured so nicely in the
following quote {\em ``In physics the only
observations we must consider are position observations...''}
\cite{Bell}.

\vskip0.5\baselineskip
{\it 2. Momentum and coordinate operators.}
 We start with the following {\it Ansatz} for the generalized momentum
operator $P$ and the corresponding ``discrete'' derivative $D$
\begin{equation}
P \ \equiv \ -i \hbar D \ =  \
- i \hbar \sum_{k=-N}^N {1\over \Delta x } \ \alpha_k
\  E^k_{\Delta x } \ ,
\label{Pdef}
\end{equation}
where the complex shift operator is defined as
\begin{equation}
E_{\Delta x }^k \ f(x) \ = \ f(x+k \Delta  x )
\ ,
\label{shiftDEF}
\end{equation}
and the shift $\Delta x$ can be complex. We keep the notation the same as in
\cite{JMP} with the exception of the overall $1/\Delta x $ factor in
the formula (\ref{Pdef}) that defines the constants $\alpha_k$'s.
With this definition our formal equations for $\alpha_k$ become
identical to those of \cite{JMP}.
The coordinate operator $X$ is assumed to have the form
\begin{equation}
X \ = \ \sum_k {1\over 2} \beta_k \ \left[
{\hat x}   E_{\Delta x}^k + E_{\Delta x}^k {\hat x}  \right]
\ ,
\label{Xdef}
\end{equation}
with unspecified constants $\beta_k$.

 We would like to have the standard coordinate (${\hat x}$) and momentum
(${\hat p }= -i\hbar \partial /\partial x$) in the continuum limit.
Hence, we impose the {\em classical limit} conditions
\begin{equation}
\lim_{\Delta x\to0} X \ = \ {\hat x}  \ ,  \qquad
\lim_{\Delta x\to0} P \ = \ {\hat p }\ .
\label{CL}
\end{equation}
In addition, we demand the symmetry:
\begin{equation}
\alpha_{-k} \ = \  \alpha_k
\label{symmetric}
\end{equation}
and that $D$ be the best fit to
$\partial /\partial x$, {\it i.e.} we assume
the operator $D$ to be {\it optimal} in the sense of \cite{JMP}.
This determines the coefficients $\alpha_k$ in a unique way, independent
of $\Delta x$
\begin{equation}
\alpha_0 \ = \  0 \ , \qquad
\alpha_k \ = \ (-1)^{k+1} {(N!)^2 \over k (N+k)! (N-k)!}
\ .
\label{alphas}
\end{equation}
Finally, the operator $X$ will be determined from
the Heisenberg CR
\begin{equation}
\left[ X, \ P \right] \ = \ i \hbar \ \ONE
\ .
\label{HeisenbergCR}
\end{equation}

 At the formal level (\ref{HeisenbergCR}) imply the same solutions
for $\beta_k$'s as in the case of the real shifts \cite{JMP}.
However, to determine Hermicity we have to specify the scalar product
and the corresponding Hilbert space. This is an especially pressing
issue for the imaginary shifts where the standard setup involving functions in
$L^2(\R)$ is not suitable.
For the real shifts the analysis of (\ref{HeisenbergCR}) has been done in \cite{JMP}.
For general complex shifts $\Delta x$ we were unable to construct
a Hilbert space on which both $X$ and $P$ would be Hermitean.
We are therefore concentrating on two types of shifts: (a) real shifts, (b) purely
imaginary shifts.

By abuse of notation we will write $\Delta x$ for real shifts, $i\Delta x$
for imaginary shifts, respectively.  Thus $\Delta x$ is always real.
The following section is devoted to a construction of a suitable
Hilbert space supporting imaginary shift representations.

\vskip0.5\baselineskip
{\it 3. Analytic setup for imaginary shifts.}
 We choose a particular model of a Hilbert space. Suppose
$\Lambda >0$ and consider a linear space
\begin{equation}
C_\Lambda = \left\{ f\in L^2(\R), \ f(p)=0 \
\hbox{\rm for } \vert p\vert >\Lambda \right\} \ .
\label{CLambda}
\end{equation}
One might think of $\Lambda$ as a momentum cut--off and $C_\Lambda$ is a
space of wave functions compactly supported in the momentum space
({\it e.g.} we could fix $\Lambda$ to be of the order of $1/l_P$, where
$l_P$ is the Plank length). In fact, we assume that
$\Lambda \sim 1/\Delta x$. It is easy to see that $C_\Lambda$ is a
closed subspace of $L^2(\R)$. Using the Fourier transform we
obtain the coordinate representation of
$C_\Lambda$ which will be denoted by $H_\Lambda$. Following \cite{Branges} we adopt the
following
\newtheorem{entire}{Definition}
\begin{entire} An entire function $F(x)$ is said to be of
exponential type at most $\Lambda$ if
$
\limsup_{\vert x\vert\to\infty} {\ln \vert F(x)\vert \over \vert x\vert}
\le \Lambda \ .
$
\end{entire}
Now, one can give the following description of $H_\Lambda$
(\cite{Branges}, Theorems 16 and 17)
\newtheorem{PW}{Theorem}
\begin{PW}[Paley--Wiener]

$H_\Lambda = \left\{ F\in L^2(\R):
F \ \hbox{\rm is entire of exponential
type at most } \Lambda \right\}$
\end{PW}
The shift operator $E_{\Delta z}$ acts on $H_\Lambda$ and, consequently,
it acts on $C_\Lambda$. Using the convention
\begin{equation}
F(x) \ = \ {1\over\sqrt{2\pi}} \int^{+\infty}_{-\infty}
e^{-ipx} \ f(p) \ dp
\ ,
\label{fourierF}
\end{equation}
we get that
\begin{equation}
E_{\Delta z}: \ f(p) \longmapsto e^{ip\Delta z} f(p) \ ,
\quad f\in C_\Lambda \ .
\label{Edz}
\end{equation}
Since $f$ vanishes outside of a compact set, $E_{\Delta z}$ is
unitary if $\Delta z = \Delta x \in \R$, Hermitean if
$\Delta z = i \Delta x$ respectively.
 The former case is analyzed in \cite{JMP} by a slightly different
method.
In this paper we are interested in the latter case. $D_{i\Delta x}$ acts
on $C_\Lambda$ as a multiplication operator by the function
\begin{equation}
D_N(p) \ = \ {1\over i \Delta x} \sum_{k=-N}^{N} \alpha_k \ z^k
\ ,
\label{DNdef}
\end{equation}
where $z=e^{\Delta x p}$.
For real $\alpha_k$, $D_N$ is skew--Hermitean. In particular, this is
true for the optimal discretization.

 Now, we turn to the study of the operator of multiplication by $x$.
We first introduce a dense subspace of
$C_\Lambda$ defined as
\[
C^0_\Lambda = \Big\{ f\in C_\Lambda : \hbox{\rm
f is a. c. } ,\
f(\Lambda) = f(-\Lambda)=0 \Big\}
\ .
\]
The corresponding subspace of $H_\Lambda$ will be denoted by
$H_\Lambda^0$.
We note that $C_\Lambda^0$ is invariant under $E_{\Delta z}$.
On $H^0_\Lambda$ we can integrate by parts obtaining
\[
xF(x) = {1\over\sqrt{2\pi}} \int_{-\Lambda}^\Lambda
e^{-ipx} {1\over i} {d\over dp} f(p) \ dp
\ .
\]
We are interested in finding pairs $X, D_N$ of operators satisfying
Heisenberg CR in the sense of definition given in
Sec. VI of \cite{JMP}.  We look for $X$ of the form
\begin{equation}
X \ = \ {1\over 2i} \ \sum_m \left\{
\beta_m \ z^m \ {d\over dp} + {d\over dp} \ \beta_m \ z^m
\right\}
\ .
\label{XAnsatz}
\end{equation}
Then the Heisenberg CR
(\ref{HeisenbergCR}) reads $[D_N(p), X] = 1$ and it implies
\begin{equation}
-{\Delta x\over i} z \left( \sum_m\beta_m z^m\right)
{d\over dz} \left(D_N(p)\right) = 1
\ .
\label{xxx1}
\end{equation}

\noindent{\it Example} For the optimal discretization scheme and $N=1$
we have found in \cite{JMP} that $\alpha_1=1/2=\alpha_{-1}$.
Then (\ref{DNdef}) implies
\[
D_1(p) = {1\over i \Delta x} \sinh (p\Delta x) \ ,
\quad -\Lambda \le p \le \Lambda \ ,
\]
and (\ref{xxx1})  yields
\[
\sum_m \beta_m z^m = {1\over (\cosh p\Delta x)} \ ,
\quad -\Lambda \le p \le \Lambda \ .
\]
In summary
\[
D_1(p) \ = \ {1\over i\Delta x} \sinh (p\Delta x) \ ,
\]
\[
X(p) \ = \ {1\over 2i} \left[ {1\over\cosh(p\Delta x)} {d\over dp}
+ {d\over dp} {1\over \cosh(p\Delta x)} \right] \ ,
\]
is a conjugate pair in the sense of \cite{Dorfmeisters}.

 Below we will show that the pair $X(p)$, $D_1(p)$ is unitary equivalent
to the canonical conjugate pair ${1\over i}{d\over dy}$, ${1\over i}y$ on
$L^2\left(\left[
{-\sinh\Lambda\Delta x\over\Delta x}, {\sinh\Lambda\Delta x\over\Delta x}
\right]\right)$.
This, in particular, implies that the spectrum of a self--adjoint
extension of $X(p)$ is a 1--dimensional lattice. It is perhaps worth
mentioning that we do not have in this case a doubling of canonical
conjugate pairs, a phenomenon occurring for real shifts \cite{JMP}.
Finally, we would like to point out that the above pair $X(p), \ D_1(p)$
appears in \cite{Frappat}. We believe that the present work provides a
proper analytic setup for that work.
For example, to define the position operator, the authors of
\cite{Frappat} had to confine their attention to the zero momentum
sector of the Hilbert space (formula (7) in \cite{Frappat}).
There is no need for such an assumption in our approach.

\vskip0.5\baselineskip
{\it 4. Optimal discretization for imaginary shifts.}
 In this Section we consider the case of the optimal discretization
scheme. We limit ourselves to the most important aspects as the details
are analogous to the real case studied in  Sec. VI of \cite{JMP}.
Our goal is to understand (\ref{xxx1}), in particular, the structure of
zeros of ${d\over dz}D_N(p)\equiv D'_N$.
We immediately have
\newtheorem{Lemma1}{Lemma}
\begin{Lemma1} Let $D_N(p)$ be optimal and $N$ be odd.
Then $D'_N(p)$ has no roots
for $p\in [-\Lambda, \Lambda]$.
\end{Lemma1}

\noindent{\it Proof:} The proof is very similar to the proof of
{\it Lemma 2} in \cite{JMP}. We compute
$i\Delta x z \; d/dz D_N$ to obtain
\[
i\Delta x z {d\over dz} D_N = (-1)^{N+1}  {(N!)^2\over (2N)!}
z^{-N} \left[
(1-z)^{2N} - (-1)^N \pmatrix{2N \cr N} z^N \right]
\ .
\]
Thus $z_0$ is a zero of $D'_N$ iff it satisfies
\[
{(1-z)^{2N}\over z^N} = (-1)^N \pmatrix{2N \cr N} \ ,
\quad z = e^{p\Delta x} \ , \quad
-\Lambda \le p \le \Lambda \ .
\]
It is clear that for odd $N$ this equation has no real roots
($z>0$).
 This completes the proof.

\begin{Lemma1} If $N$ is even and $\Lambda = 1/\Delta x$, then
$D'_N(p)$ has no roots for
$p\in [-\Lambda, \Lambda]$.
\end{Lemma1}

\noindent{\it Proof:}
Now, the equation determining the zeros of $D'_N$ reads
\[
{(1-z)^{2N}\over z^N} = \pmatrix{2N \cr N} \ ,
\quad z = e^{p\Delta x} \ ,
\quad -\Lambda \le p \le \Lambda \ .
\]
In other words
\[
\sinh^2{p\Delta x\over 2} = a_N \ ,
\quad a_N ={1\over 4} ^N\sqrt{\pmatrix{2N \cr N}} \ ,
\quad -\Lambda \le p \le \Lambda \ .
\]
In \cite{JMP} we have proved that $\{a_N\}$ is increasing and
$\lim_{N\to\infty} a_N = 1$. Note that
\[
0 \le \sinh^2{p\Delta x\over 2} \le \sinh^2{\Lambda \Delta x\over 2}
\ .
\]

\noindent Thus we have no roots iff
$\sinh^2{\Lambda\Delta x\over 2} < a_N$ for every even $N$.

\noindent Since $\{a_N\}$ is increasing the latter inequality is
equivalent to
\[
\sinh^2{\Lambda\Delta x\over 2} < a_2 = {\sqrt{6}\over 4} \ ,
\]
which implies that
\[
\Lambda \Delta x < 2  \hbox{\rm arcsh} \left( {^4\sqrt{6}\over 2} \right)
\simeq 1.4379... \ ,
\]
which holds if $\Lambda = 1/\Delta x$.

\noindent{\it Remark:} It is clear from the proof of {\it Lemma 2}
that if $\Lambda\Delta x$ exceeds the bound 1.4379... then $D'_2$
has exactly two roots. This case is then very similar to the case of
real shifts (\cite{JMP}, Sec. VI).
 From now on we assume that $D'_N$ has no real roots on the interval
of interest. One easily checks if $-\Lambda \le p \le \Lambda $,
then
$i\Delta x D'_N >0$ if $N$ is odd and $i\Delta x D'_N < 0$ if $N$ is even.
Hence, the map
\[
\omega : \ p \longrightarrow iD_N(p) \ , \quad
[-\Lambda, \Lambda] \longrightarrow [ -|D_N(\Lambda)|, |D_N(\Lambda)|]
\]
is bijective. Let us point out that $iD_N(p)$ is real for
$p\in[-\Lambda,\Lambda]$. We call the new variable $y=iD_N(p)$.
The corresponding Hilbert space is
$L^2\left([-|D_N(\Lambda)|, |D_N(\Lambda)|]\right)$.
The map $\omega$ induces an isomorphism of
$L^2\left([-\Lambda, \Lambda]\right)$ and
$L^2\left([-|D_N(-\Lambda)|, |D_N(\Lambda)|]\right)$.
In particular, $\omega$ maps
\begin{equation}
X \longrightarrow {1\over i} {d\over dy} \ , \qquad
D_N \longrightarrow {1\over i} y
\ .
\label{omegamap}
\end{equation}
This situation should be contrasted with the results we have obtained in
\cite{JMP} for real shifts, for which the pairs $X, D_N$ are, after a
suitable rescaling, unitarily
equivalent to {\it two} copies of the canonical conjugate pairs ${1\over i} {d\over dy}, \,{1\over i} y$ defined on $L^2([-1,1])$.

\vskip0.5\baselineskip
{\it 5. Imaginary vs. real shifts.}
 The aim of this section is to clarify the differences between the real
and imaginary shift representations of the Heisenberg CR.
In doing so we will come across an interesting connection to problems in
signal theory, centered around the so--called Shannon theorem
\cite{Debanchis}.
Let us consider $f\in C_\Lambda$. Every such function can be written
in terms of its Fourier coefficients. We define
\begin{equation}
\hat f(n) \ = \ {1\over 2\Lambda} \int_{-\Lambda}^\Lambda
e^{-i{n\pi\over\Lambda}p} \ f(p) \ dp
\ .
\label{Fourier1}
\end{equation}
In view of (\ref{fourierF}) we obtain
\begin{equation}
F\left({n\pi\over\Lambda}\right) \ = \ {2\Lambda\over\sqrt{2\pi}}
\hat f(n) \ .
\label{Fourier2}
\end{equation}

\noindent Since
$f(p) \ = \ \sum_n \hat f(n) \ e^{i{n\pi\over\Lambda}p}$,
we conclude that

\begin{equation}
f(p) \ = \ {\sqrt{2\pi}\over 2\Lambda} \sum_n
F\left({n\pi\over\Lambda}\right) \ e^{i{n\pi\over \Lambda}p}
\ .
\label{Fourier4}
\end{equation}
Using the definition of $F(x)$ we arrive at {\it Shannon's theorem}
\begin{equation}
F(x) \ = \ \sum_n F\left( {n\pi\over\Lambda}\right) \
{\sin(n\pi-\Lambda x)\over n\pi - \Lambda x}
\ ,
\quad F \in H_\Lambda \ .
\label{ShannonTh}
\end{equation}
Thus $F(x)$ can be recovered from its sample values on the lattice
${\pi\over\Lambda} \Z$.

 Our objective now is to understand how the real and imaginary shifts
act on those sample values. It is clear that the real shift $E_{\Delta x}$, for
$\Delta x=\pi/\Lambda$, gives
$\left( E_{\Delta x}F\right)(n\pi/\Lambda)=F({n+1\over\Lambda}\pi)$.
The question that presents itself at this point is whether there exists
a simple formula for the action of the imaginary shift. We know that the
action of $E_{i\Delta x}$ is most conveniently expressed on $C_\Lambda$
by $f(p) \mapsto e^{p\Delta x} f(p)$. Thus, in view of (\ref{Fourier2})
we obtain
\[
\left( E_{i\Delta x} F \right) \left( {n\pi\over \Lambda}\right)
= {2\Lambda\over\sqrt{2\pi}} \widehat{(e^{p\Delta x} f)}(n)
\ ,
\]
and with the help of the convolution theorem and (\ref{Fourier2})
we get
\begin{equation}
\left( E_{i\Delta x} F \right) \left( {n\pi\over \Lambda}\right) =
(-1)^n \sinh(\Lambda\Delta x) \sum_{m\in\Z} (-1)^m
{F\left({m\pi\over\Lambda}\right)\over\Lambda\Delta x - i(n-m)\pi}
\ .
\label{Fourier6}
\end{equation}

 We conclude that the imaginary shift $E_{i\Delta x}$ involves
{\it infinite number of real shifts} by
$\pi/\Lambda$. One can easily extend these computations to the ``discrete''
derivative $D_{i\Delta x}$. For simplicity, we only consider here the case
$N=1$ for which we obtain
\begin{equation}
\left(D^1_{i\Delta x} F\right) \left({n\pi\over\Lambda}\right)
= (-1)^n {\sinh(\Lambda\Delta x)\over\Delta x}
\sum_{m\in\Z} (-1)^m {(m-n) \pi \over
(\Lambda \Delta x)^2 + (m-n)^2\pi^2}
F\left({m\pi\over\Lambda}\right)
\ .
\label{Fourier7}
\end{equation}
We remark that the main contribution to the sum (\ref{Fourier7})
comes form the region $m\sim n \pm {\Lambda\Delta x\over \pi}$.
Since $\Lambda \sim 1/\Delta x$, we conclude that (\ref{Fourier7})
can be interpreted as a smeared central difference scheme. However, due
to the slow decay one cannot ignore the tail of the expansion
(\ref{Fourier7}).
It is perhaps fair to say that the utility of difference schemes
involving infinitely many points is questionable from a computational
point of view. If, however, one is willing to accept that in principle
our goal is not to find useful computational methods of the existing
quantum theory but explore the foundations of quantum kinematics, then
the imaginary shifts merit a careful study.
In the next section we comment on the problem of fermion doubling viewed
from the perspective of different choices of the discrete derivative.

\vskip0.5\baselineskip
{\it 6. Remarks on fermion doubling.}
 In this Section we briefly comment on the problem of the spectrum
doubling for the lattice Dirac operator.
For an elementary introduction to the problem one can consult
\cite{Kogut}.
 Let us consider the 2--D lattice Dirac equation
\begin{equation}
i\hbar {\partial\over\partial t} \Psi \ = \ \gamma_5 \ P \ \Psi
\ ,
\label{DiracEq}
\end{equation}
where $\gamma_5=\pmatrix{0 & 1 \cr 1 & 0}$, $P$ is a momentum
operator.
If one uses, say, the central difference scheme
\[
{\partial\over \partial x} \Psi(x) \longmapsto
{\Psi(n\Delta x + \Delta x) - \Psi(n\Delta x - \Delta x)\over 2\Delta x}
\ ,
\]
followed by the plane wave substitution
\[
\Psi(n\Delta x) \ = \ e^{-i\omega t+i n \Delta x p} \Phi(n)
\ ,
\]
one obtains the dispersion relation
\[
\omega^2 \ = \ {\sin^2(p\Delta x)\over(\Delta x)^2}
\ .
\]
Spectrum doubling arises because the right hand side has two zeros in
the Brillouin zone. We have shown in \cite{JMP} that the spectrum doubling
is tantamount to the presence of two copies of the canonical
conjugate pairs.  This effect occurs for any optimal discretization and real
$\Delta x$.  In particular,
the central difference scheme is included in that scheme and corresponds
to $N=1$.
The situation for the imaginary shift is different!
In the simplest case of $N=1$ we have
\[
P \ = \ {1\over  \Delta x} \ \sinh(p\Delta x)
\ , \quad -\Lambda \le p \le \Lambda \ ,
\]
and the dispersion relation reads
\[
\omega^2 \ = \ {\sinh^2 (p\Delta x) \over \Delta x^2} \ ,
\quad  -\Lambda \le p \le \Lambda \ .
\]
There is no spectrum doubling; only a single copy of the canonical conjugate
pair occurs. One can directly interpret (\ref{DiracEq}) as a lattice
equation by using the interpretation of the imaginary shift in terms of
infinitely many real shifts presented in Sec. V.

\vskip0.5\baselineskip
{\it 7. Summary and conclusions.}
This paper concludes the project of looking for Heisenberg
algebra representations in terms of the complex shift operators
of the form (\ref{Pdef}) and (\ref{Xdef}). In the present paper
we have limited our
attention to the Schr\"odinger picture, hence to the Hermitean $X,P$
pairs. For the case of purely imaginary shifts we have formulated a Hilbert space approach using the space of
entire functions.
Assuming furthermore the {\it optimal} form of the momentum operator introduced in
\cite{JMP} we have found that
{\it (i)} the spectrum of the
self--adjoint extension of the coordinate operator $X$ comprises a
lattice,
{\it (ii)} all the conjugate pairs are, after a suitable
rescaling, unitary equivalent to the canonical pair ${1\over i}y, \, {1\over i}{d\over dy}$ on $L^2([-1,1])$
{\it (iii)} the pairs $X,D$ do not exhibit the
doubling which occurs for real shifts.

 As the next step within this approach one could analyze evolution
(wave) equations for state vectors and operators corresponding to
observables other than $X,P$.
Some work in this direction has been done  a long time ago
(\cite{Pais,Das,Cole}).  We, nevertheless, expect that the most interesting
feature arising from our work, which is the lattice structure of the
spectrum of $X$, has not been sufficiently elucidated.  In particular,
one could pursue a potentially interesting analogy with the solid state
physics. But this will be a subject of the future research.

\acknowledgments
 The authors gratefully acknowledge the support of the State Committee
for Scientific Research (KBN) grant no 2 P03B 140 10
and the NSERC grant no OGP0138591.

\end{document}